\documentclass{raa01}            
\usepackage{graphicx,times}
\usepackage{natbib}
\usepackage{lscape}
\bibpunct{(}{)}{;}{a}{}{,}

\providecommand{\bm}[1]{\mbox{\boldmath$#1$\unboldmath}}

\begin{document}
\title{Extragalactic dispersion measures of fast radio bursts}

\volnopage{{\bf 2015} Vol.\ {\bf 15} No. {\bf 10}, 1629--1638~
{\small doi: 10.1088/1674--4527/15/10/002}}
      \setcounter{page}{1629}

   \author{Jun Xu
   \and J. L. Han
   }

\institute{
National Astronomical Observatories, Chinese Academy of
  Sciences, Beijing 100012, China; {\it xujun@nao.cas.cn}\\
\vs \no
{\small Received 2014 December 11; accepted 2015 April 1}}

\abstract{  Fast radio bursts show large dispersion measures, much
larger than the Galactic dispersion measure foreground. Therefore,
they evidently have an extragalactic origin.
We investigate possible contributions to the dispersion measure
from host galaxies.
We simulate the spatial distribution of fast radio bursts and calculate the
dispersion measures along the sightlines from fast radio bursts
to the edge of host galaxies by using the scaled NE2001 model for
thermal electron density distributions.
We find that contributions to the dispersion measure of fast radio
bursts from the host galaxy follow a skew Gaussian distribution. The
peak and the width at half maximum of the dispersion measure
distribution increase with the inclination angle of a spiral galaxy,
to large values when the inclination angle is over $70\degr$. The
largest dispersion measure produced by an edge-on spiral galaxy can
reach a few thousand pc~cm$^{-3}$, while the dispersion measures
from dwarf galaxies and elliptical galaxies have a maximum of only a
few tens of pc~cm$^{-3}$.
Notice, however, that additional dispersion measures of tens to
hundreds of pc~cm$^{-3}$ can be produced by high density clumps
in host galaxies.
Simulations that include dispersion measure contributions from the
Large Magellanic Cloud and the Andromeda Galaxy are shown as
examples to demonstrate how to extract the dispersion measure from
the intergalactic medium.
}

\authorrunning{J. Xu \& J. L. Han }    
\titlerunning{Extragalactic DMs of FRBs}
   \maketitle


%

\section{Introduction}
\label{sect:intro}

In the Milky Way, measurements of dispersion measures (DMs) of pulsars
have been used to delineate the electron density distribution of the
interstellar medium (ISM). Outside our Galaxy, only 21 extragalactic
pulsars in the Magellanic Clouds have DM measurements
\citep{mfl+06}. In recent years, pulsar surveys have discovered a new
category of short-duration (a few milliseconds) radio bursts which are
called fast radio bursts (FRBs). Due to the short duration of FRBs,
the scale sizes of such exotic astrophysical events must be
small. Possible origins of FRBs have been proposed, such as black hole
evaporation \citep{kskl12}, coalescence of neutron stars
\citep{tot13}, binary white-dwarf mergers \citep{kim13}, supermassive
rotating neutron stars collapsing to black holes \citep{fr14},
synchrotron maser emission from relativistic magnetized shocks
\citep{lyu14}, magnetar hyperflares \citep{pp10}, and nearby flaring
stars \citep{lsm14}. Maybe some FRBs are related to gamma-ray bursts
(GRBs) \citep[e.g.][]{zha14}.  To date, ten such events have been
reported \citep{lbm+07,kskl12,tsb+13,sch+14,bb14,pbb+14,rsj15}. Most
of these FRBs are found at high Galactic latitudes, and have large DMs
up to several hundreds pc~cm$^{-3}$, exceeding the possible Galactic
foreground DM. Therefore FRBs are believed to have an extragalactic
origin.

DMs of a large number of pulsars or FRBs in external galaxies can be
used to detect the electron density distribution in the diffuse
intergalactic medium (IGM) and ISM in the host galaxies
\citep[e.g.][]{den14,zok+14,hvl+14}.  The DM of a given short duration
source at distance ($D$, in units of pc) is defined as the integral of
the free electron density ($n_{\rm e}$, in units of cm$^{-3}$) along
the line of sight
\begin{equation}
\mathrm{DM}=\int_0^D n_{\rm e} dl.
\label{dm}
\end{equation}
The observed DM, $\mathrm{DM}_{\rm obs}$, of an extragalactic object
is the sum of the DM from the host galaxy, $\mathrm{DM}_{\rm host}$,
DM in intergalactic space, $\mathrm{DM}_{\rm IGM}$, and the foreground
DM contributed by our Galaxy, $\mathrm{DM}_{\rm Gal}$, i.e.
\begin{equation}
\mathrm{DM}_{\rm obs}=\mathrm{DM}_{\rm host} + \mathrm{DM}_{\rm IGM} +
\mathrm{DM}_{\rm Gal}.
\label{dmobs}
\end{equation}
The Galactic foreground DM is the integral of electron density along
the line of sight from the Sun to the furthest reaches of the Milky
Way, which can be estimated by different models of free electron
distribution as discussed by \citet{sch12}. The popular thermal
electron density model for the Milky Way is NE2001 \citep{cl02} which
is mostly built upon pulsar DM measurements.  However, all known
pulsars, except for 21 pulsars in the Large Magellanic Cloud (LMC) and
the Small Magellanic Cloud, are located inside the Milky
Way. Therefore, the electron density distribution beyond pulsars but
in the extended Galactic halo cannot be well measured at present, and
hence the uncertainty of the NE2001 model in the halo region cannot be
assessed. Little is known about the DMs contributed by external
galaxies. \citet{lbm+07} estimated a probability of 25\% for the DM
contribution of 100 pc~cm$^{-3}$ from a host galaxy. \citet{tsb+13}
discussed the DM contribution of host galaxies and estimated
$\mathrm{DM}_{\rm host} \approx $ 100 pc~cm$^{-3}$ in a spiral galaxy
with a median inclination angle of 60$\degr$. The rough distance or
redshift estimation of FRBs relies on $\mathrm{DM}_{\rm
  IGM}=\mathrm{DM}_{\rm obs} - \mathrm{DM}_{\rm Gal} -
\mathrm{DM}_{\rm host}$ \citep[see][]{tsb+13,lbm+07}, with the
assumption of $\mathrm{DM}_{\rm host}$ = 100 pc~cm$^{-3}$. Although
the observed DMs in high-redshift galaxies can be small due to
cosmological time dilation and the associated frequency shift
\citep{zlw+14}, the local DM contribution from host galaxies to FRBs
is very important for calculating the DM from the IGM and then
estimating their cosmological distances.

Currently, the origin of FRBs is not known, and the types and
morphologies of host galaxies are also poorly constrained. In this
paper, we apply Monte Carlo simulations to the spatial distribution of
FRBs in a galaxy, and model the probability distribution of DM from
different types of host galaxies. The NE2001 model is scaled according
to the integrated H$\alpha$ intensity to represent the electron
density distributions in host galaxies. We simulate the DM
contribution from spiral galaxies in Section 2, and discuss DM
distributions in other types of galaxies in Section 3. In Section 4 we
discuss the DM caused by clumps. In Section 5 a specific simulation
for the LMC and the Andromeda Galaxy (M31) is provided. The
conclusions are given in Section 6.

\section{DM contribution from spiral galaxies}
\label{simulation}

Most FRBs probably occur in spiral galaxies with locations
corresponding to stars or related objects. Here we model their spatial
distribution and show their probable DM distributions. To specify the
locations of any modeled FRBs, we use a Cartesian coordinate system
($x$, $y$, $z$) where the center of the host galaxy is at the
origin. The spatial distribution of FRBs in a galaxy is described by
an exponential function for height and a Gaussian radial
distribution. The surface density of FRBs is therefore likely to be
large near the galactic center, and decreases with radial
distance. The scale height of the two-sided exponential function is
set to be 330 pc, and the characteristic radius for the Gaussian
radial density profile is set to be 6.5 kpc, similar to that for the
simulation of a pulsar population \citep{lfl+06}.



For the Milky Way, the popular model for the density distribution of
free electrons is NE2001 \citep{cl02}. The Galactic DM along a line of
sight from the boundary of the Milky Way to the Sun can be calculated
through the model. For other galaxies the DMs from a host galaxy can
be calculated along a sightline for the integrals of free electron
density from a specific location to the near boundary of the
galaxy. However, we know little about the electron density
distributions in other galaxies. For a general spiral galaxy, the size
and gas density structure are not very different from the Milky Way
\citep{bf15}. We therefore model the electron density distribution
in a spiral galaxy (size of $\sim$ 30 kpc) using the thin disk, thick
disk, spiral arm and Galactic Center components of the NE2001 model
but ignoring small-scale clumps, voids and the local ISM.


\begin{figure}
\centering
\includegraphics[height=60mm,angle=-90]{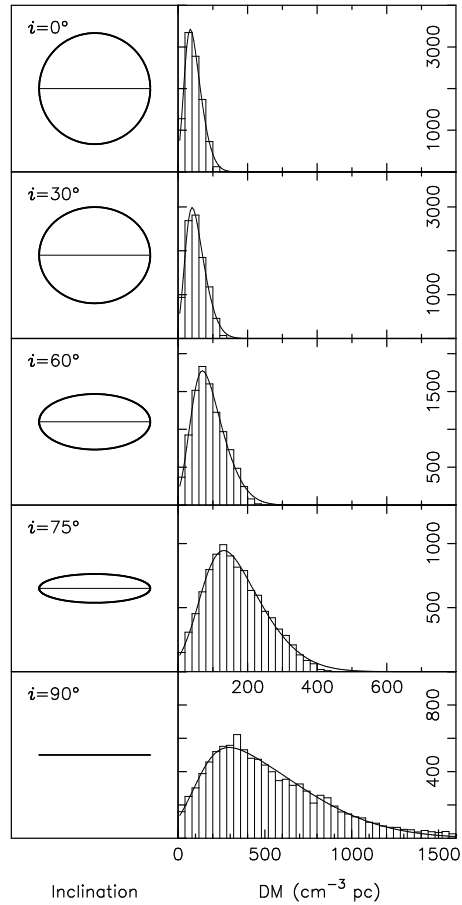}

\caption{\baselineskip 3.6mm DM distributions of FRBs in a spiral
galaxy with different inclination angles (0$\degr$, 30$\degr$,
60$\degr$, 75$\degr$ and 90$\degr$). The solid lines represent the
fittings to the distribution with a skew Gaussian function.}
\label{DMplot}
\end{figure}

\begin{table}

\caption{\baselineskip 3.6mm Fitting parameters of a skew Gaussian
  function for the DM distributions of FRBs in spiral galaxies for ten
  inclination angles, with the peak DM (column 6) and values for the
  left and right half width at half maximum $W_{\rm DM-L}$ and $W_{\rm
    DM-R}$.} \label{tab}

\centering

\fns\tabcolsep 3mm
\begin{tabular}{l c c c c c c c}
\hline\noalign{\smallskip}
  $i$  & $\xi$ & $\omega$   &  $\alpha$ & $W_{\rm DM-L}$ & DM$_{\rm peak}$ & $W_{\rm DM-R}$   & DM$_{\rm median}$  \\
 (deg)& \multicolumn{2}{c}{({pc~}cm$^{-3}$)}& &\multicolumn{3}{c}{({pc~}cm$^{-3}$)} &({pc~}cm$^{-3}$)\\
  (1)  & (2)  & (3)  & (4)  & (5)  & (6)  & (7)  & (8)  \\
\noalign{\smallskip}\hline\noalign{\smallskip}

   0   & 16.7 &  38.9  &  2.9  & 21.6  & 35.3    & 33.3   & 42.5   \\
   10  & 16.9 &  39.8  &  3.0  & 21.8  & 35.7    & 34.2   & 43.1   \\
   20  & 17.6 &  41.9  &  3.0  & 22.7  & 37.3    & 36.0   & 45.2   \\
   30  & 19.3 &  45.1  &  3.1  & 24.3  & 40.4    & 38.6   & 49.0   \\
   40  & 23.0 &  49.4  &  2.7  & 28.9  & 47.4    & 42.2   & 55.3   \\
   50  & 27.3 &  58.5  &  2.9  & 32.7  & 55.4    & 50.2   & 65.8   \\
   60  & 34.4 &  75.3  &  3.0  & 41.6  & 70.3    & 64.4   & 84.0   \\
   70  & 49.3 &  108.3 &  2.9  & 60.6  & 101.3   & 92.8   &121.1   \\
   80  & 86.2 &  198.8 &  2.7  & 116.0 & 184.1   & 170.6  &219.5   \\
   90  & 92.7 &  527.4 &  4.8  & 215.3 & 293.8   & 466.7  &455.2   \\

\noalign{\smallskip}\hline
\end{tabular}
\end{table}

\begin{figure}
\centering
\includegraphics[height=60mm,angle=-90]{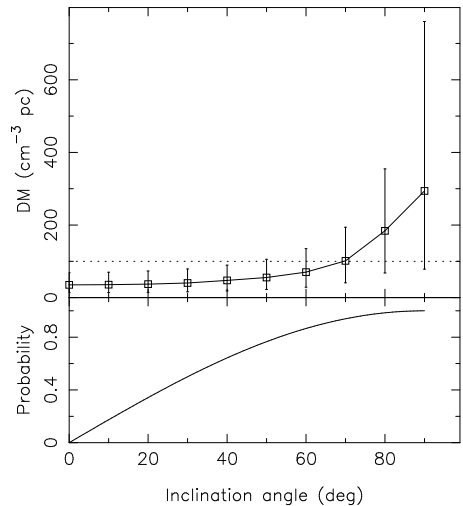}

\caption{\baselineskip 3.6mm The peak DM ({\it squares}) and the
width at half maximum ({\it the vertical bars}) for FRBs in spiral
galaxies change with the inclination angle $i$. The dotted line is
100 pc~cm$^{-3}$. The lower panel shows the probability distribution
of the inclination angles oriented randomly in space, which should
follow the curve $\sin(i)$.} \label{sgDMi}
\end{figure}

With this basic model for the electron density distribution, we use a
Monte Carlo simulation to generate the spatial distribution of FRBs in
a spiral galaxy for {10\,000} locations. When observed FRBs come from
a distant galaxy outside the Milky Way, all sightlines to these FRBs
are toward us and nearly parallel. Each spiral galaxy has a specific
inclination angle, $i$, with respect to the sightline. Obviously,
FRBs behind the midplane of the disk should show a larger DM than
those in front of the midplane. We calculate the DM of each FRB by
integrating the free electron density in the host galaxy along the
line of sight from its location to the nearer boundary at 15 kpc from
the galactic center. Figure~\ref{DMplot} shows the DM distributions
for {10\,000} assumed FRBs in a spiral galaxy with different
inclination angles (0$\degr$, 30$\degr$, 60$\degr$, 75$\degr$ and
90$\degr$). The DMs show a peak in the range of 30 to 300
pc~cm$^{-3}$, and shift to a large value with increasing inclination
angles. The distribution follows a skew Gaussian function, defined by
\begin{equation}
\frac{dN}{d\mathrm{DM}}=N_0~e^{-\frac{(\mathrm{DM}-\xi)^2}{2\omega^2}}~\int_{-\infty}^{\alpha(\frac{\mathrm{DM}-\xi}{\omega})}e^{-\frac{t^2}{2}}dt,
\label{sgdis}
\end{equation}
where $\frac{dN}{d\mathrm{DM}}$ is the number per DM, and $N_0$ is the
normalization constant; $\xi$, $\omega$ and $\alpha$ are the location,
scale and shape parameters, respectively; $t$ is the variable used in
the integral function. The results of fitting parameters for the DM
distributions are listed in Table~\ref{tab}.  The peak DM, DM$_{\rm
  peak}$, and the left and right half width at half maximum, $W_{\rm
  DM-L}$ and $W_{\rm DM-R}$, are given in column{s} (5) -- (7)
respectively. The peak DMs and the widths at half maximum are
plotted in Figure~\ref{sgDMi} as a function of the inclination angle
$i$, both of which increase very quickly at large inclination
angles. The probability distribution of the inclination angles of
spiral galaxies follows the function $\sin(i)$, as shown in the
lower panel in Figure~\ref{sgDMi}. The peak DM exceeds 100
pc~cm$^{-3}$ when the inclination angle is over 70$\degr$. In a
face-on galaxy, when the line of sight passes vertically through the
whole disk, the DM reaches the maximum of a few tens of
pc~cm$^{-3}$, except for the central regions that have a size of
$\sim$100 pc where the DM can be a few hundred pc~cm$^{-3}$
\citep{dcl09}. When the host galaxy is edge-on, the sightline to an
FRB is at a low height and the very farthest side passes through the
entire thin disk, so that the DM can be up to 4670 pc~cm$^{-3}$. For
the peak DM and the width at half maximum, we find empirical
formula{e} to describe
$\mathrm{DM}_\mathrm{peak}\sim33.8+0.94e^{0.06i}$,
$W_\mathrm{DM-L}\sim22.1+0.21e^{0.08i}$, and
$W_\mathrm{DM-R}\sim39.5+0.02e^{0.11i}$. We calculate the ensemble
distribution marginalized over the inclination angle and derive the
weighted average of the DM distribution for a spiral galaxy to be
142 pc~cm$^{-3}$. If the FRBs follow the distribution of stars with
a scale height of young Galactic Population I objects ($\sim$ 100
pc) \citep{gr83}, the DM distribution will be more symmetric and
have a narrower distribution, but the DM peaks at the same value as
shown by the simulations above.

\section{DM contribution from other types of galaxies}

FRBs may also happen in dwarf galaxies or elliptical galaxies.
The electron density distributions of the two types of galaxies
are unknown. However, according to their shapes and structures,
we can model their electron density distributions by scaling
the NE2001 model according to the total intensity of H$\alpha$
emission because H$\alpha$ is a tracer of
ionized gas. The emission measure (EM) is the pathlength integral
of electron density squared, i.e. $\mathrm{EM}=\int{n_{\rm e}^2dl}$.
The total H$\alpha$ luminosity of a galaxy is proportional
to electron density squared in the volume.

Based on the galactic structures and their total H$\alpha$
luminosity, we construct the electron density model for dwarf
galaxies and elliptical galaxies, and study their DM contribution.
Dwarf galaxies include irregular galaxies, dwarf spirals and dwarf
ellipticals. Here, we are concentrated on young and gas-rich
Magellanic type galaxies, which are typical dwarf galaxies. The size
of Magellanic type dwarfs is about 1 -- 5 kpc and the H$\alpha$
luminosity is 10$^{37}$ -- 10$^{39}$ erg~s$^{-1}$
\citep{jsb+04,kkr12}. Magellanic type dwarfs are typical irregular
galaxies, but they have unconspicuous spiral arms and disks. For
simplicity, we assume that the size of a dwarf is 3 kpc and the
H$\alpha$ luminosity is 10$^{38}$ erg~s$^{-1}$. The total H$\alpha$
luminosity of the Milky Way is about 10$^{40}$ erg~s$^{-1}$, as
inferred from statistics of Milky Way-like galaxies \citep{jsb+04}.
We scale the H$\alpha$ luminosity of the Milky Way from a diameter
of 30 kpc to the size of a dwarf (3 kpc), and set the H$\alpha$
luminosity to 10$^{37}$ erg~s$^{-1}$, which is 1/10 of the observed
value from a dwarf. The electron density model of a dwarf is
therefore obtained by scaling the physical size to 1/10 and the
amplitude to $\sqrt{10}$ those of the NE2001 model.

\begin{figure}[hpt!]
\centering
\includegraphics[height=70mm,angle=-90]{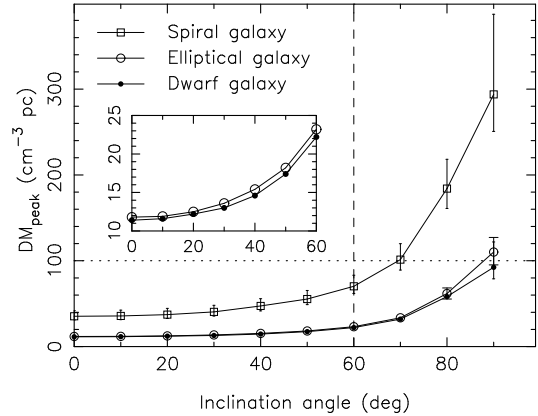}

\vspace{-3mm}\caption{\baselineskip 3.6mm The peak value of DM
distributions for three types of galaxies as a function of the
inclination angle $i$. The width at half maximum is plotted in the
vertical bar (scaled to 1/5 of its original value for clarity). The
dotted line is plotted at the value of 100 pc~cm$^{-3}$. The
dashed line marks the half probability of the inclination angle
distribution.} \label{galDMi}
\end{figure}

Elliptical galaxies vary greatly in size, from 10 kpc to over 100
kpc. There is very little ISM in elliptical galaxies, which results
in low rates of star formation. The H$\alpha$ luminosity of a common
elliptical galaxy is 10$^{38}$ -- 10$^{40}$ erg~s$^{-1}$
\citep{ksc+14}. The shapes of elliptical galaxies are similar to the
bulges of spiral galaxies. Here we consider a normal elliptical
galaxy with the same size as the Milky Way, and
its H$\alpha$ luminosity is 10$^{39}$ erg~s$^{-1}$. So, in the
modeling of electron density in an elliptical galaxy, we only use
the thick disk and Galactic Center components of the NE2001 model.
We then scale the electron density components by reducing the
amplitude to $\sqrt{1/10}$ according to the H$\alpha$ luminosity.
The inclination angle $i$ is defined by the angle between the line of
sight and normal direction of the largest plane with the major axis.

In the Monte Carlo simulations, we scale the spatial distribution of
{10\,000} FRBs within a galaxy accordingly. The DM distributions of
FRBs in these galaxies can also be fitted with a skew Gaussian
function as well, but the peak DM and the width at half maximum are
much smaller than the corresponding values for spiral galaxies, as
shown in Figure~\ref{galDMi}. FRBs from dwarf galaxies and
elliptical galaxies have small DMs of a few to a few tens of
pc~cm$^{-3}$. The DMs increase slowly from 11 -- 12 pc~cm$^{-3}$ at
$i = 0\degr$  to 22 -- 24 pc~cm$^{-3}$ at $i = 60\degr$, and may
reach 100 pc~cm$^{-3}$ when $i \sim 90\degr$. The ensemble average
of the DM distribution for a dwarf galaxy is 45 pc~cm$^{-3}$, and
for an elliptical galaxy it is 37 pc~cm$^{-3}$.

\section{DM caused by high density regions in a galaxy}

In the previous sections, only a large-scale diffuse ISM was
considered during simulations for DM contributions from host
galaxies. It is worth noting that there are thousands of small-scale
high density regions in a galaxy like the Milky Way, where the
electron density is strongly enhanced relative to the ambient
density of the ISM. Though their volume filling factor is quite
small (with the maximum of a few tens of pc), the sightline toward
an FRB to pass through a clump would produce significant DM.

Clumps of ionized gas most likely correspond to discrete
H\,\textsc{ii} regions. To date, more than 2500 H\,\textsc{ii} regions
have been observed in our Galaxy \citep{hh14}.  Their distribution is
very inhomogeneous and most of them are concentrated in spiral
arms. Extragalactic H\,\textsc{ii} regions are also found in spiral
galaxies and irregular galaxies, but very few are detected in
elliptical galaxies \citep{zlw+14}. The size of a Galactic
H\,\textsc{ii} region ranges from a few percent of a pc (ultracompact)
to several tens of pc, which is roughly inversely proportional to the
electron density in an H\,\textsc{ii} region.  The typical size and
average electron density of H\,\textsc{ii} regions detected by the
Sino-German 6 cm Polarization Survey
\citep[e.g.][]{shr+07,grh+10,srh+11,xhr+11} are a few tens of pc and
a few cm$^{-3}$, respectively. The DMs caused by the clumps range from
tens of pc~cm$^{-3}$ to hundreds of pc~cm$^{-3}$ according to
previous studies of pulsars in our Galaxy. Therefore, if an FRB in
the host galaxy happens to take place behind a clump, a large part of
the observed DM could come from the clump.

\section{Applications in the LMC and M31}

In the near future, a next generation radio telescope, the Square
Kilometre Array (SKA), will reveal the electron density distribution
in nearby galaxies and the IGM \citep{hvl+14}. DM measurements of
extragalactic pulsars and single pulses are the first powerful probes
that can be used to detect the extragalactic medium, including ISM in
a host galaxy and IGM. A large sample of pulsars or FRBs in nearby
galaxies such as the LMC and M31 will be detected by this
highly-sensitive telescope \citep[see][]{kbk+15}. A large sample of
Galactic pulsars in the region of sky immediately around the direction
of the host galaxy can be used to estimate the foreground column
density of electrons in the Milky Way. After discounting the
foreground DM contribution from our Milky Way and constraining the
local contribution from the host galaxy, the DM from IGM can be
derived from observations of extragalactic pulses. Here, we apply the
simulations for host galaxies of the LMC and M31, to demonstrate the
detection of intergalactic DM.

\begin{figure}[t!]
\centering
\includegraphics[width=120mm,angle=0]{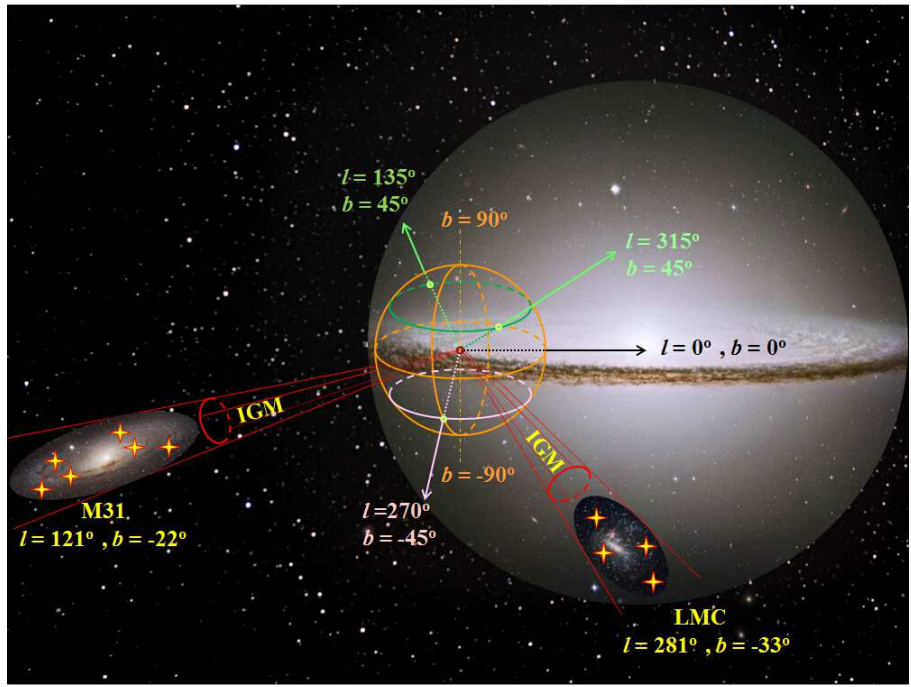}

\caption{\baselineskip 3.6mm A schematic diagram showing the locations of the
LMC and M31 in Galactic coordinates.} \label{Schem}
\end{figure}

\begin{figure}[t!]
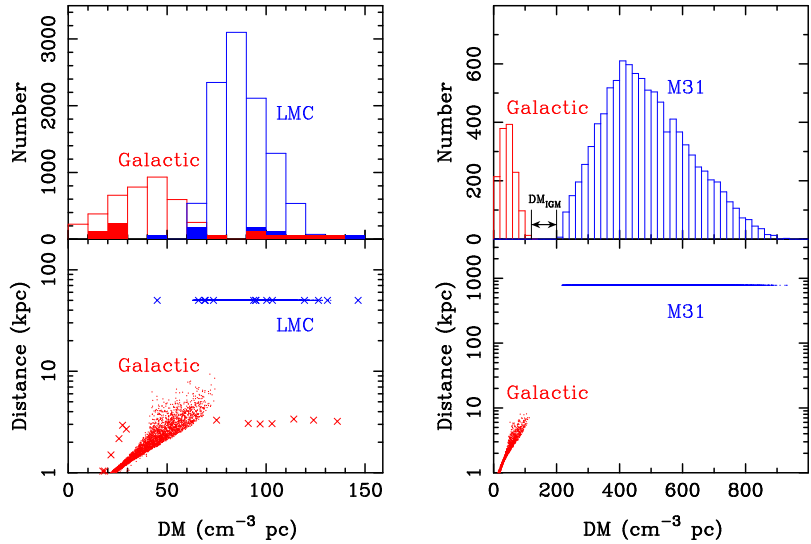

\centering
\includegraphics[height=50mm,angle=-90]{ms2107fig5a.ps} \hspace{5mm}
\includegraphics[height=50mm,angle=-90]{ms2107fig5b.ps}

\caption{\baselineskip 3.6mm Simulated DM distributions of
  extragalactic pulsars (blue) and Galactic pulsars (red) within a
  10$\degr$ radius around the LMC {\it on the left} and M31 {\it on
    the right}.  A small sample of pulsars in the LMC and a small
  sample of foreground Galactic pulsars around the cloud are marked
  with crosses ({\it lower left panel}) and presented with as the {\it
    scaled} filled area of the histogram ({\it upper left
    panel}). Currently the NE2001 model of electron density
  distribution is not good enough for reconstructing the DMs of
  foreground pulsar data. However, in the future, SKA will discover a
  much larger sample of Galactic pulsars to improve the electron
  density distribution model and discover many extragalactic pulsars
  (e.g.\ in M31) that can be used to deduce the DM contribution from
  host galaxies and the IGM.
\label{IGM}}
\end{figure}

The LMC is a satellite of our Milky Way, at a distance of about 50 kpc
\citep{pgg+13} in the southern hemisphere of the sky ({$l=280.5\degr$,
  $b=-32.9\degr$}). Its size is about 4.3 kpc, with an inclination of
$\sim$35$\degr$ \citep[e.g.][]{ndk+04}. At present, 15 pulsars have
been discovered in the LMC, and their DMs have been estimated for 13
pulsars \citep{mfl+06}. M31 is the largest galaxy in the Local Group,
with a distance of approximately 780 kpc \citep{sg98} in the direction
({$l=121.2\degr$, $b=-21.6\degr$}). M31 is estimated to have an
inclination angle of 77$\degr$ \citep[e.g.][]{ccf09}. The apparent
diameter is about 40 kpc. Their locations relative to the Milky Way
are illustrated in Figure~\ref{Schem}. The simulations mentioned in
Sections 2 and 3 are scaled for the LMC and M31, respectively, as
shown in Figure~\ref{IGM}. We simulate foreground Galactic pulsars
within a 10$\degr$ radius around the directions of host galaxies. The
``observed'' DMs from the host galaxies in Figure~\ref{IGM} are the
sum of DM from the host galaxies, DM in the IGM and the foreground DM
contributed by our Galaxy in their directions. We add
DM$_\mathrm{IGM}$ of 10 pc~cm$^{-3}$ for the LMC, and 150 pc~cm$^{-3}$
for M31 according to their distance.

Current data for pulsars in the LMC and in the Milky Way near the
direction of LMC are compared with a simulated pulse distribution in
the left panel of Figure~\ref{IGM}. Obviously, the current NE2001
model of electron density distribution is not good enough to
reconstruct the DMs of farthest Galactic pulsars. It is expected that
an improved electron density distribution model can be constructed in
the future. The simulated DMs for pulsars in the LMC are very
consistent with the observed DMs of LMC pulsars, except for two
outliers.

As pointed out by \citet{hvl+14}, it is difficult to estimate the
dispersion contributed by the host galaxy for only an individual
extragalactic pulsar or pulse. A large sample of pulsars is required
for this purpose. As shown in the right panel of Figure~\ref{IGM}, the
minimum DM of these pulsars, $\min(\mathrm{DM}_\mathrm{extraPSRs})$,
can be expressed as the upper limit of DM$_\mathrm{IGM}$ plus the
foreground Galactic DM, $\max(\mathrm{DM}_\mathrm{GalacPSRs})$. The
electron density of the IGM can be estimated by
$\mathrm{DM}_\mathrm{IGM}=
\min(\mathrm{DM}_\mathrm{extraPSRs})-\max(\mathrm{DM}_\mathrm{GalacPSRs})$.
When large samples of pulsars or FRBs in a host galaxy and also in the
Galactic region in the general direction of a host galaxy are
discovered by SKA in the future, the electron density distribution
model for the Milky Way will be greatly improved, and the DM
contribution from the host galaxy and the IGM can then be deduced.

\section{Conclusions}

We apply Monte Carlo simulations to generate the spatial distribution
of FRBs in a host galaxy. The electron density distribution of the
host galaxy is modeled by using the scaled model of NE2001. The DM
distributions of FRBs from the host galaxies are calculated for
different types of galaxies, which roughly follow a skew Gaussian
function. The characteristic parameters of the skew Gaussian function
increase with inclination angle. FRBs or pulsars in spiral galaxies
can have a large DM when the inclination angle of the galaxy is over
$70\degr$ and the sightline of an FRB passes through the central
region. The largest DM can reach a few thousand pc~cm$^{-3}$ when a
galaxy is edge-on. The possible DM distributions of pulsars or FRBs in
dwarf galaxies and elliptical galaxies are in the range of a few to a
few tens of pc~cm$^{-3}$. In addition, high density clumps will induce
a large DM for pulsars or FRBs of tens to hundreds of pc~cm$^{-3}$ if
they are behind the clumps.

In the future when many FRBs or extragalactic pulsars are detected,
the statistics of their DM distribution can be used to derive the DM
contribution by the host galaxy and derive the intergalactic DMs after
discounting the foreground Galactic DMs.

\normalem

\begin{acknowledgements}
The authors are supported by the National Natural Science Foundation
of China (Grant Nos.~11473034 and 11503038), and the Strategic
Priority Research Program ``The Emergence of Cosmological
Structures'' of the Chinese Academy of Sciences (Grant
No.~XDB09010200).
\end{acknowledgements}

\label{lastpage}
\end{document}